# Filamentary Surface Plasma Discharge Flow Length and Time Scales

Lalit K. Rajendran[1*], Bhavini Singh[1], Pavlos P. Vlachos[2], and Sally P.M. Bane[1]

[1] Purdue University, School of Aeronautics and Astronautics, West Lafayette, USA.

[2] Purdue University, School of Mechanical Engineering, West Lafayette, USA.

*lrajendr@purdue.edu

## Abstract

Nanosecond Surface Dielectric Barrier Discharge (ns-SDBDs) are a class of plasma actuators that utilize a high-voltage pulse of nanosecond duration between two surface-mounted electrodes to create an electrical breakdown of air, along with rapid heating. These actuators usually produce multiple filaments when operated at high pulse frequencies, and the rapid heating leads to the formation of shock waves and complex flow fields. In this work we replicate a single filament of the ns-SDBDs and characterize the induced flow using velocity measurements from particle image velocimetry and density measurements from background-oriented schlieren. The discharge is produced by a high voltage electrical pulse between two copper electrodes on an acrylic base. A hot gas kernel characterizes the flow field formed close to the electrodes that expands and cools over time and a vortex ring that propagates away from the surface while entraining cold ambient fluid. The gas density deficit inside the kernel displays a power-law decay over time. Based on the observations, we develop a simplified theoretical model based on vortex-driven cooling and perform a scaling analysis to obtain the induced flow length and time scales. The results show that the cooling process's time scales correspond to a circulation-based time scale of the vortex ring, and the length scale of the kernel corresponds to the vortex ring radius. These findings can guide the choice of optimal filament spacing and pulse frequencies in the design, deployment, and operation of nanosecond surface dielectric barrier discharges (ns-SDBDs) for flow control.

# 1   Introduction

There is growing interest in the use of nanosecond surface dielectric barrier discharge (ns-SDBD) actuators for high-speed (supersonic/hypersonic) flow control. A plasma discharge is created in these actuators using a nanosecond-duration pulse of several kilovolts to deposit energy rapidly in the electrode gap [1]–[3]. The electrical breakdown leads to a two-step ultra-fast heating mechanism characterized by (1) the electronic excitation of gaseous nitrogen molecules by electron impact and (2) subsequent dissociative quenching of the excited $N_2$ by oxygen molecules producing oxygen atoms and excess thermal energy [4]–[6]. The rapid heat release leads to the formation of a shock wave and the development of a complex three-dimensional flow field near the actuator surface characterized by coherent vorticity and a hot gas kernel [7]–[13].

Past work studying the flow induced by ns-SDBDs has shown that the initial strength of the induced shock increases with the peak voltage and that the shock rapidly decays to an acoustic wave on moving away from the actuator surface [7], [8], [14], [15]. High-speed schlieren visualization of the post-shock stage of the induced flow has shown the presence of a hot gas kernel near the electrodes, which expands and cools to ambient [11], [13], [15], [16]. Actuators based on ns-SDBDs have also been applied to control the shock-boundary layer interaction (SBLI) on a wedge, and it was found that the actuator can perturb the low-frequency unsteadiness in the separation bubble [17]. Interactions between the shock generated by the actuator and the incident oblique shock have also been observed in the study by Kinefuchi et al. [18], [19], who found an optimal pulse frequency for the actuator, corresponding to a time scale based on the boundary layer thickness and the flow velocity. Although a general idea of the flow features induced by a ns-SDBD exists, the effect of the actuator geometry (such as the filament spacing) and the operating parameters (such as the pulse frequency) on the induced flow are not well understood and play a critical role in flow control applications. Further, a knowledge of the intrinsic frequency of the flow induced by the actuator and its relation to the time scales associated with an oncoming flow is also critical for flow control applications.

Even the flow field induced by a single pulse of a ns-SDBD is not entirely understood at a more fundamental level, in contrast to the well-characterized AC-driven SDBD. The flow field induced by ns-DBDs is on much shorter time scales (by almost an order of magnitude) and involves large spatiotemporal gradients in the velocity and temperature fields, posing a significant experimental challenge. However, detailed measurements of the induced flow are required to develop a mechanistic model of the actuator performance, such as the vorticity production, heating, penetration depth, etc., and to develop scaling rules that relate the actuator design and operating conditions to its performance.

In this work, we perform the first detailed characterization of the flow induced by a single filamentary surface discharge produced by a single nanosecond-duration pulse in a quiescent medium using Particle Image Velocimetry (PIV) and Background Oriented Schlieren (BOS) measurements and develop a reduced-order model for the flow field. Filamentary discharges are preferred because they can provide localized heating with minimal power density requirements and provide better control authority as their position on the surface and morphology is known and controllable [10], [11], [13]. Further, Leonov et al. [13] have observed that when ns-SDBD actuators are operated in high Reynolds number flows, they almost always transform to a filamentary discharge due to ionization instabilities. While reducing the problem to a single filament and a single pulse is a considerable simplification from practical applications, it allows us to remove the interaction between the flow induced by adjacent filaments and subsequent pulses.

We first identify a candidate actuator that can be used to create a well-controlled single plasma filament with a single pulse and then perform PIV and BOS measurements to characterize the induced flow for a range of discharge energies. The measurements show that the induced flow consists of a hot gas kernel filled with vorticity in a vortex ring that expands and cools over time. We also develop a reduced-order model to describe the induced flow and show that the expansion of the kernel is governed by the vortex ring motion, and the entrainment of cold gas governs the cooling. Applying the model to the experimental data reveals that the vortex ring's properties govern the time scale associated with the kernel dynamics. The model predictions for the actuator-induced flow length and time scales can guide the choice of filament spacing and pulse frequencies for practical multi-pulse ns-SDBD configurations.

## 2 Experimental Methods

### 2.1 Actuator Geometry, Plasma Generation and Electrical Measurements

A saw-tooth actuator consisting of copper electrodes on an acrylic base was used for the measurements and is shown in Figure 1 (a). The electrodes tips were 2 mm apart, and the thickness of the copper tape was approximately 0.1 mm. This particular actuator was chosen because it could produce a single discharge filament for a single pulse, thereby providing a controlled line deposition of energy, as shown in Figure 1 (b). These choices were based on direct imaging of the discharge produced by several actuator designs with varying tip geometries, and with some candidates also featuring Kapton tape to provide higher dielectric strength. Since the candidates with Kapton tape could not produce a single filament with a single pulse, which is the main objective of this study, this particular actuator was chosen even though it is not a traditional DBD. The actuator designs were based on previous work by Devarajan et al. [11], and the details on the design study are summarized in [20].

The plasma discharge was generated by an Eagle Harbor Technology NSP-300 high voltage nanosecond pulser connected to the electrodes by high voltage cables soldered to the electrodes. The electrical properties of the discharge were measured using two Tektronix P6015A high voltage probes (in a differential measurement configuration) and a Magnelab CT-D1.0 current transformer. The probes were connected to an Agilent DSO9104A oscilloscope, and the voltage and current traces were used to calculate the discharge power and energy deposited. The displacement current was measured for a case without a breakdown to calculate the lag between the voltage and the current [6]. The energy was varied by changing the DC voltage and the pulse duration to obtain a range of 1 – 5 mJ.

## 2.2 Background-Oriented Schlieren (BOS)

Background-Oriented Schlieren (BOS) was used to measure the density of the hot gas kernel and to characterize the cooling process [21]–[27]. A schematic of the experimental setup is shown in Figure 1 (c). The induced flow was imaged perpendicular to the electrode axis (in the Y-Z plane in Figure 1). The dot pattern used was a photomask with a regular grid of dots manufactured by FrontRange Photomask with dot diameters of 42 µm and center-to-center dot spacing of 42 µm. The photomask was back-illuminated by an arc lamp, and a diffuser plate was placed between the lamp and the dot pattern to provide uniform illumination over the field of view. The images were acquired using a Photron SA-Z camera recording at 20 kHz at 1024 x1024 pixels. The camera was equipped with a 200 mm focal length Nikon lens and a 2X teleconverter and operated at an aperture stop of f16 to achieve a magnification of 10 $\mu$m/pixels and a field of view of 10 mm x 10 mm. The distance between the dot pattern and the density gradients ($Z_D$) was 25.4 mm, and the distance between the camera and the dot pattern ($Z_B = Z_D + Z_A$) was 267 mm. A sample magnified image of the dot pattern at the first time instant with the plasma filament is shown in Figure 1 (b). It should be noted that this particular image is not used for the displacement estimation, and the image analysis begins from the 2nd frame.

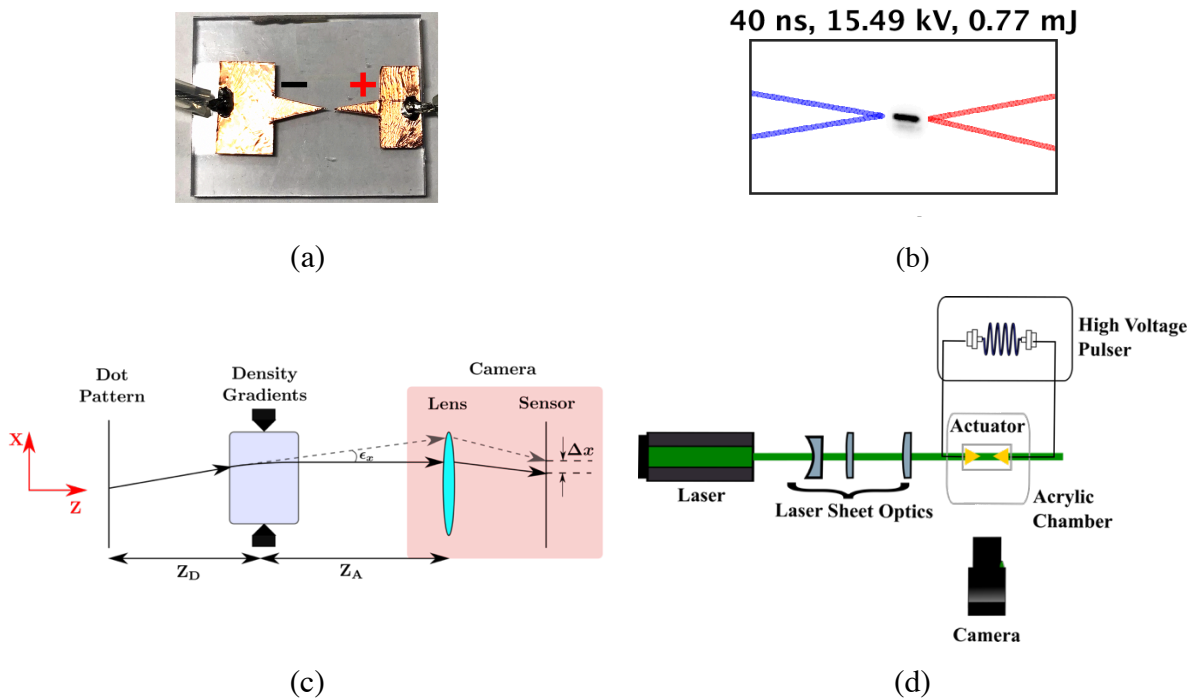

**Figure 1.** (a) Saw-tooth actuator used for the experiments, (b) false-color image of the discharge filament, and (c) schematic of the BOS experimental setup (top view), and (d) schematic of PIV experimental setup (top view).

The dot pattern images with and without the plasma-induced flow were processed using a dot tracking methodology that has been shown to provide an order of magnitude improvements in the accuracy, precision, and spatial resolution of the displacement estimation for BOS [28]. The method utilizes prior information about the dot pattern design, such as the location, size, and the number of dots, to provide near 100% yield. A correlation correction is performed after the tracking to improve the dynamic range for subpixel displacement estimation. The displacement estimates were validated with a Universal Outlier Detection (UOD) method for unstructured measurements using Delaunay triangulation [29]. The displacement uncertainties were calculated using a recently developed methodology for uncertainty amplification in BOS, based on the ratio of the cross-correlation plane diameters of the dot intensity maps in the reference and gradient images [30].

The displacement fields were used to calculate the density gradients using Equation (1),

$$\frac{\partial \rho_p}{\partial x} = \int \frac{\partial \rho}{\partial x} dz = \frac{\Delta x}{Z_D M} \frac{n_0}{K} \quad (1)$$

where $\partial \rho / \partial x$ is the density gradient along the $x$ direction, $\Delta x$ is the pixel displacement on the camera sensor, $M$ is the magnification, $Z_D$ is the distance between the density gradient and the dot pattern, $n_0$ is the refractive index of the undisturbed medium, and $K$ is the Gladstone-Dale constant. The displacements were interpolated onto a regular grid (with the grid spacing based on the target dot spacing) along with the uncertainties, and a 2D integration was performed using a Weighted Least Squares density integration procedure [31] to obtain the projected density field $\rho_p$ *relative* to the ambient [32]. Dirichlet boundary conditions were used on the left and right boundaries with zero relative projected density, as these points correspond to the ambient. The final spatial resolution of the measurements was 0.08 mm. The displacement uncertainties were propagated through the density integration procedure to estimate the density uncertainty [33], and the maximum uncertainties in the projected density were about 3% of the peak density deficit with respect to the ambient.

Following the density integration, a *projected density deficit* $\rho_d = \rho_{p,\infty} - \rho_p$ was calculated where $\rho_{p,\infty} = 0$, and the hot gas kernel was identified as the set of points with a projected density deficit greater than 5% of the peak density deficit. Then the mean density deficit of all points in the kernel was calculated for each time instant, and this procedure was performed at each time step for all tests.

## 2.3 Particle Image Velocimetry (PIV)

Time-resolved planar Particle Image Velocimetry (PIV) was used to measure the velocity and vorticity fields induced by the surface discharge. A schematic of the PIV system is shown in Figure 1 (d). The setup consists of an enclosed acrylic test section containing the surface discharge actuator, a Photron SA-Z camera, and an EdgeWave Nd:YAG laser operating at 20 kHz. The laser sheet optics produced an approximately 1 mm thin waist in the region of interest where the plasma was generated. A Quantum Composer Model 575 delay generator was used to synchronize and trigger the laser, cameras, and high voltage pulse generator. A fluidized bed seeder was used to inject aluminum oxide particles with diameters of about 0.3 $\mu$m and estimated Stokes number of approximately 0.002 into the chamber. Particle images were recorded at 20,000 fps at a resolution of 1024 x 1024 pixels using the Photron camber with a Nikon Nikorr 105 mm lens.

PRANA (PIV Research and ANAlysis) software was used to process the recorded particle images [34]. The correlation method used was the Robust Phase Correlation (RPC) [35]–[37] in an iterative multigrid framework using window deformation [38]–[41], with each pass validated by universal outlier detection (UOD) [42]. A total of four passes was used, and a 50% Gaussian window was applied to the original window size [36], resulting in window resolutions of 64 x 64 pixels in the first pass to 32 x 32 pixels in the last pass, with 50% window overlap in all passes. Between successive passes, velocity interpolation was performed using bicubic interpolation, and the image interpolation was performed using a sinc interpolation with a Blackman filter. The subpixel displacement was estimated using a three-point Gaussian fit [43], and the displacement uncertainty was calculated using the Moment of Correlation method [44]. The measurements' final spatial resolution was 0.16 mm, and the average uncertainty was approximately 0.02 m/s, about 20% of the mean velocity.

The velocity measurements were de-noised based on Proper Orthogonal Decomposition (POD) before post-processing [45]. The vorticity was calculated from the velocity field using the 4$^{th}$ order noise-optimized compact-Richardson scheme [46]. As the vorticity calculations cannot differentiate between shear and swirl regions, coherent structures (vortex cores) identification was performed using a $\lambda_{CI}$ criterion, or the swirl strength [47]. Regions in the flow field characterized by a swirl greater than the instantaneous 95$^{th}$ percentile were considered coherent/swirling. The vorticity in the other regions was set to zero to ensure that spurious and shear-based vorticity measurements did not affect the subsequent calculation of the vortex ring properties.

The vortex ring parameters, such as the circulation and ring radius, were then calculated based on integral relations [48]. Instead of explicit tracking of the vortex cores, this method was used to minimize errors induced by coherent structure identification due to the complex distribution of vorticity. Further, the integral relations derived initially for an axisymmetric vortex ring were modified to account for the general experimental case, which may violate these assumptions because of tilting, three-dimensionality, etc. Therefore, the *net circulation* $\Gamma$ was defined as half of the area integral of the magnitude of the vorticity field as given by (2), and the ring centroid was defined as the first moment of the magnitude of the vorticity field as given by (3). Finally, the *ring radius R* was defined as the first moment of the magnitude of the vorticity distribution about the ring centroid and is given by (4).

$$\Gamma = \frac{1}{2} \iint |\omega_z(x,y)| \, dx \, dy \qquad (2)$$

$$x_0 = \frac{\iint x\, |\omega_z(x,y)|\, dxdy}{\iint |\omega_z(x,y)|\, dxdy} \quad (3)$$

$$R = \frac{\iint |x - x_0|\, |\omega_z(x,y)|\, dxdy}{\iint |\omega_z(x,y)|\, dxdy} \quad (4)$$

These definitions reduce to the standard definition of the vortex ring properties for a perfectly axisymmetric flow field [48].

## 3 Results

In this section, we present observations of the density and vorticity fields to show that the discharge induces a hot gas kernel and a vortex ring, both of which move away from the surface over time. To describe the cooling process and the vortex dynamics, we develop a model relating the time variation of the mean kernel density to the vortex ring properties and use the model to develop characteristic length and time scales of the induced flow. The results presented in this work correspond to measurements on an actuator with an electrode gap of 2 mm, viewed from the side, obtained from a set of 15 tests for PIV, and another separate set of 15 tests for BOS.

### 3.1 Measurements of the plasma-induced flow field

The density and velocity measurements of the flow field from a single test are presented in Figure 2, though it should be noted that the two measurements are obtained from different realizations of the induced flow. The density measurements show a torus-shaped hot gas kernel close to the discharge location. Over time, the kernel propagates upwards, expands, and cools. In addition, the vorticity measurements show the formation of a vortex ring that propagates upwards from the surface. The ring is seen to entrain ambient fluid from the side and ejects this fluid vertically away from the surface.

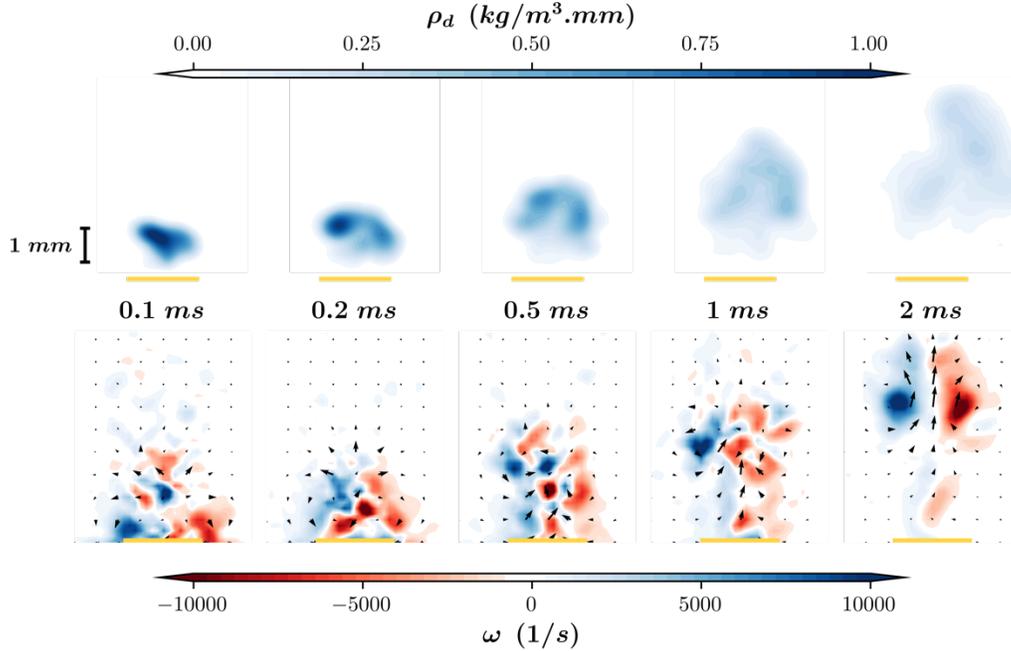

**Figure 2.** Density and vorticity fields induced by the surface discharge from 0.1 to 2 ms. The top row shows the projected density deficit, and the bottom row shows the vorticity contours and velocity vectors. The two sets of measurements correspond to different realizations of the induced flow.

The time histories of the bulk properties of the hot gas kernel and the vortex ring are shown in Figure 3 (a) and (b), respectively. Figure 3(a) shows that the mean density deficit of the kernel increases at early times, followed by a cooling period, and the simultaneous increase in area denotes the kernel expansion. Both the density deficits and the kernel areas increase with energy deposited in the plasma. For the vortex ring properties, shown in Figure 3(b), there is no such apparent effect of energy deposited, and the time series of the measurements is quite noisy.

The results show that the density deficit of the hot gas kernel decreases with time, and since the cooling process is essentially a surrogate for passive scalar mixing, we are interested in the time scale of this process and its relation to other flow parameters. In the case of a filamentary discharge produced between two pointed electrodes far away from a surface, the cooling rate of the hot gas kernel is controlled by cold gas entrainment due to a pair of vortex rings induced near the electrode tips [49], [50]. As the current flow induced by the surface discharge also features a vortex ring and a hot gas kernel, we are interested to know if a similar coupling exists between them. We examine these issues in the next section by reducing a simplified model for the induced flow.

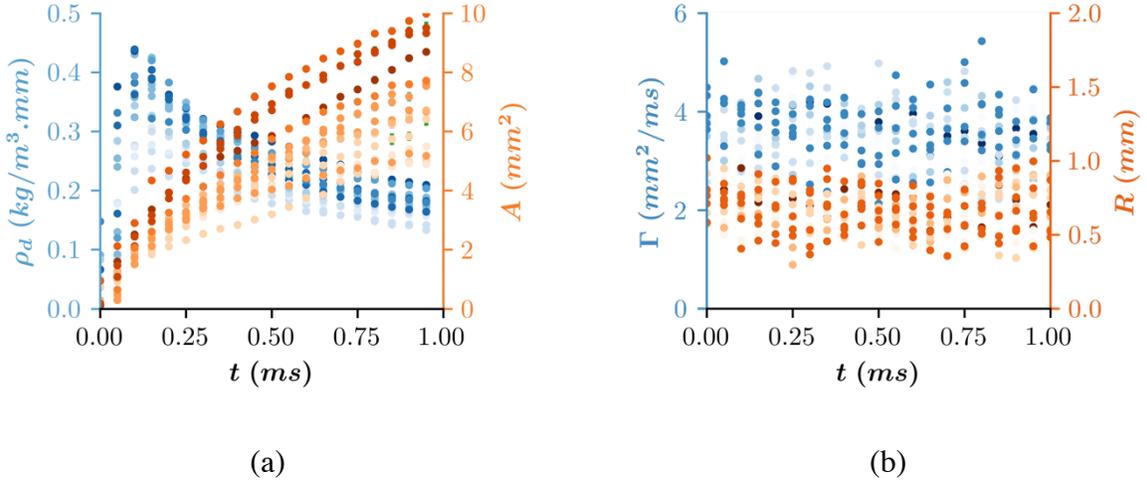

**Figure 3**. Time histories of kernel and ring properties. (a) Density deficit of the kernel (left axis) and kernel area (right axis). (b) Net circulation (left axis) and radius (right axis) of the vortex ring. The darker markers represent a case with higher energy deposited.

## 3.2  A simplified model for the induced flow

This section develops a reduced-order model for the cooling induced by the vortex ring and performs a scaling analysis. A sketch of the flow field is shown in Figure 4 with a hot gas kernel near the wall and a vortex ring inside the kernel. The kernel is modeled as a cylindrical control volume denoted by the gray dashed line in the figure, and it is further assumed that this cylinder expands purely along the vertical direction due to the ring motion. The vortex ring entrains cold fluid along the hot gas kernel's sides and ejects warm, mixed fluid through the top boundary. Under this framework, and following the analysis of Singh. et al. [49], [50], the relation between the density of the hot gas kernel and the entrainment can be expressed by,

$$\frac{\rho_\infty - \rho_k}{\rho_\infty - \rho_{k,i}} = \exp\left(-\int_{t_i}^{t} \frac{\dot{V}_{in}}{V_k} dt\right) \qquad (5)$$

where $\rho_\infty$ is the ambient density, $\rho_k$ is the mean kernel density, $V_k$ is the kernel volume, $\dot{V}_{in}$ is the entrainment of gas into the kernel, and the subscript $i$ represents the initial conditions. Equation (5) is derived by combing the inviscid mass and energy conservation equations in the low Mach number limit. Details of the derivation are given in [49].

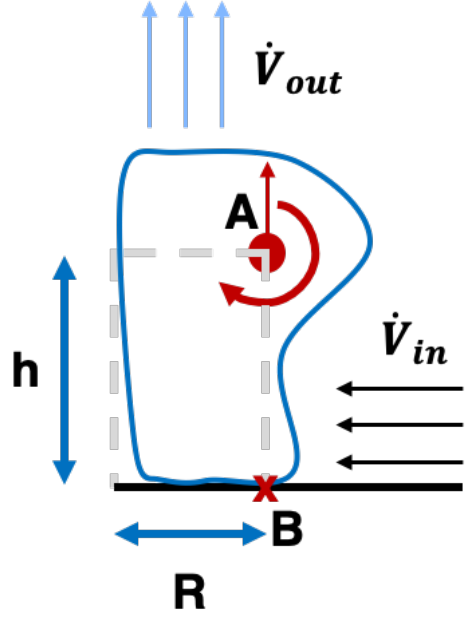

**Figure 4.** Schematic of a simple model for the flow field induced by the surface filament discharge.

In this analysis, we evaluate Equation (5) for the present geometry by modeling the vortex ring as a thin-core ring with a uniform vorticity distribution [51] and simplify the equation based on a scaling analysis. This flow field has an added feature due to the presence of the wall, which can potentially affect the vortex ring dynamics and the cooling process. As shown by Walker et. al. [52], a vortex ring can be affected by a nearby wall due to the (1) no-penetration (inviscid), and (2) no-slip (viscous) boundary conditions. The inviscid effect is accounted for by modeling an image vortex below the wall, affecting properties such as the ring diameter, entrainment, etc. The viscous effects influence the velocity profile in the boundary layer near the wall and the shear stress distribution. Walker et al. noted from both computations and experiments that both the effect of the image vortex and viscous effects are negligible when the distance of the primary vortex from the wall is greater than one ring radius. In this situation, the entrainment can then be ascribed purely to the primary vortex ring, and the contribution from wall effects is negligible. This entrainment through the right side of the control volume can be expressed as the difference between the value of the streamfunction $\psi$ at the core of the ring (point A) and at the wall (point B). Further, since the streamfunction due to a vortex ring decays rapidly away from the core, its value at B when h > R will be negligible. The entrainment can then be expressed as

$$\dot{V}_{in} = 2\pi(\psi_A - \psi_B) \qquad (6)$$
$$= \Gamma R \left[ \log\left(\frac{8R}{a}\right) - \frac{3}{2} \right] \approx \Gamma R.$$

where $\Gamma$ is the net circulation, $R$ is the ring radius, and $a$ is the vortex core radius. The volume $V_k$ in Equation (5) can be calculated from the cylindrical control volume properties, which expands due to the ring motion. When the distance between the ring and the wall is larger than the ring radius, $h > R$, the ring can be approximated to have constant properties, and the volume of the ring as a function of time can be expressed as,

$$V_k(t) = \pi R^2 h(t) \qquad (7)$$
$$= \pi R^2 V_{R,y} t$$
$$= \frac{\pi R^2 \Gamma}{2\pi R} t \approx \Gamma R t$$

where $V_{R,y}$ is ring velocity along the vertical direction. Under these assumptions, the cooling equation in (5) can be simplified to obtain

$$\frac{\rho_\infty - \rho_k}{\rho_\infty - \rho_{k,i}} = \frac{\rho_{d,k}}{\rho_{d,k_i}} \qquad (8)$$
$$= \exp\left(-\int_{t_i}^t \frac{\alpha}{t} dt\right) = \left(\frac{t}{t_i}\right)^{-\alpha}$$

where $\rho_{d,k}$ is the average density deficit for the points within the hot gas kernel, $\rho_{d,k_i}$ is the kernel deficit at the start of the cooling process (taken to be the maximum across the time series), and $\alpha$ is a constant of proportionality in the scaling analysis. The time $t_i$ is the initial condition for the cooling analysis, and since the analysis is only valid once $h > R$, $t_i$ is taken as the time instant corresponding to when the vortex ring is one radius away from the wall ($h = R$). This time $t_i$ can be related to the vortex ring properties, and it can be shown to that $t_i = \frac{R^2}{\Gamma} = \tau$.

From this simple model, we obtain the result that (1) the cooling of the hot gas kernel is described by a power-law process over time, and (2) the time scale of this process is determined by the vortex ring properties such as the circulation and ring radius. In the next section, we compare these model results to the experimental measurements.

## 3.3 Length and time scale analysis

To test the model result that the cooling follows a power law process, we replot Figure 3 (a) with a normalized density deficit and time scale. The density deficit normalized by the peak value $\left(\frac{\rho_{d,k}}{\rho_{d,k,0}}\right)$ and the time is normalized based on the time scale obtained from a power-law fit of equation (8) to the raw measurements. This is shown in Figure 5 (a), and we observe a near-collapse of the density deficit time series from all tests, thereby showing that a universal power-law process indeed describes the cooling of the kernel. Next, we also compare the time scale from the power-law fit $\tau_k$ to that from the vortex ring properties ($\tau_r = R^2/\Gamma$) in Figure 5 (b), and observe a close agreement between the two estimates ($\sim 0.1 - 0.2$ ms), thereby showing that there is indeed an effect of the vortex ring on the cooling process. The variability in the results may be because the two measurements correspond to different realizations of the induced flow.

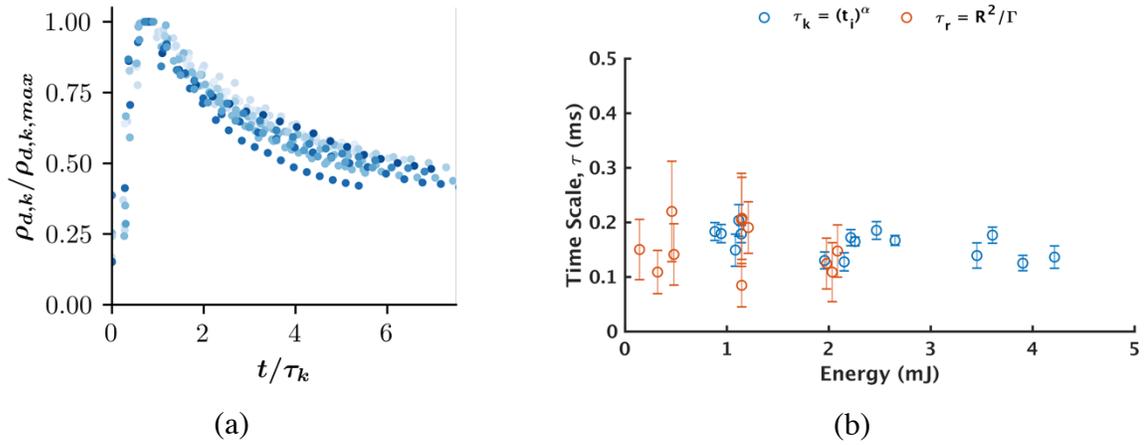

**Figure 5.** (a) Time history of the normalized density deficit shows a near-collapse of all cooling curves. (b) Comparison of time scales obtained from the kernel and ring properties.

Next, we also compare the length scale of the hot gas kernel to the vortex ring radius. Figure 6 (a) shows the time history of the kernel area (previously shown in Figure 3 (a)) normalized by its mean value across time, and we again see a collapse of the area curves across all tests. In Figure 6 (b), we compare a representative length of the kernel based on the time average area $\lambda_k = \frac{1}{2}\sqrt{A_k}$, (with the factor of two to account for the assumed axisymmetry of the model) to the vortex ring radius, and again observe that both estimates take similar values over the energy range from about $0.5 - 1$ mm.

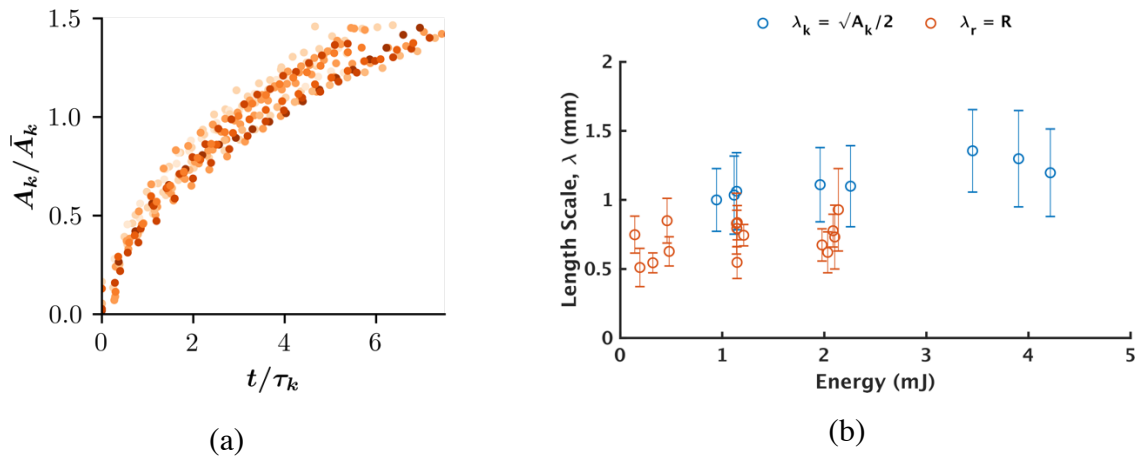

**Figure 6.** (a) Time history of the normalized kernel area. (b) Comparison of length scales from the kernel properties to the ring radius.

# 4 Conclusions

The results of this work show that a vortex ring-based cooling model is able to predict the length and time scales of the flow induced by a nanosecond surface discharge. This is consistent with recent findings that vortex rings drive entrainment and cooling in flow induced by pin-to-pin spark discharges [50]. While the present analysis is about the cooling of the hot gas kernel, the fluid density is just one example of a passive scalar, and the results can be generalized to the mixing of any quantity in the induced flow. For example, in a situation where the actuator is placed in the presence of an oncoming flow, the vortex ring may also introduce mixing of momentum. In the case of a high-speed chemically reacting flow, the actuator may result in the mixing of chemical species in addition to temperature.

The model predictions of the inherent length and time scales of the induced flow also provide useful guidelines on the optimal spacing and pulse frequency for a multi-filament, multi-pulse ns-SDBD actuator, which are more commonly used in flow control applications. For example, to ensure optimal mixing, the spacing of the plasma filaments is to be one vortex ring diameter, and the optimal frequency/time interval between pulses should be based on the vortex ring circulation and diameter. As the decay of the kernel follows a power-law variation, one can specify the time interval required for a specified decay/mixing percentage (say 50%), and use this percentage to determine the optimal inter-pulse interval. This is because, in a power-law process, the rate of mixing reduces with time, so it may be more beneficial to pulse the actuator repeatedly after 50% mixing, with longer wait times yielding diminishing returns.

One of the unanswered questions in this work is the origin of the vorticity and formation of the vortex ring. It is possible that the vorticity may be controlled by the properties of the shock wave (such as curvature and speed) that is observed at early times [53]. Suppose a model for predicting this vorticity is available, and the formation mechanism of the vortex ring is understood. In that case, one can use the analysis in this work to estimate the desired vortex ring properties for a given operating condition (such as filament spacing and pulse frequency) and estimate the required energy and electrode gap. On the other hand, if there is a specified power budget, one can estimate the vortex ring properties and then design the filament configuration. Such a model for the vortex ring formation would also help connect the early and late stages of the induced flow.

Finally, it was observed during the experimental campaign that the flow was highly three-dimensional. Volumetric measurements of the velocity and density fields using a tomographic PIV and BOS measurement system are required to investigate the flow's three-dimensionality further. This might also help characterize the orientation of the vorticity and heat flux transport and their effect on flow control, as reported by Kinefuchi et al. for SBLI [18], where they observed that the orientation of the actuator (filaments) with respect to the free-stream affected the size of the separation bubble, which they hypothesized a due to competing effects of heat and vorticity generation by the plasma.

## 5 Acknowledgment

This material is based upon work supported by the U.S. Department of Energy, Office of Science, Office of Fusion Energy Sciences under Award Number DE-SC0018156. Ravichandra Jagannath and Nick Schmidt are acknowledged for help with the actuator design and conducting the experiments.